\documentclass[aps,prb,twocolumn,superscriptaddress,showpacs]{revtex4-1}

\usepackage{graphicx}
\usepackage{calc}
\usepackage{bm}
\usepackage{color}
\usepackage{textcomp}

\begin{document}

\title{Non-linear Hall effect in three-dimensional Weyl and Dirac semimetals}

\author{O.O.~Shvetsov}
\author{V.D.~Esin}
\author{A.V.~Timonina}
\author{N.N.~Kolesnikov}
\author{E.V.~Deviatov}
\affiliation{Institute of Solid State Physics of the Russian Academy of Sciences, Chernogolovka, Moscow District, 2 Academician Ossipyan str., 142432 Russia}

\date{\today}

\begin{abstract}
    We experimentally investigate a non-linear Hall effect  for three-dimensional  WTe$_2$ and  Cd$_3$As$_2$ single crystals, representing  Weyl and Dirac  semimetals, respectively. We observe finite second-harmonic Hall voltage, which  depends quadratically on the longitudinal current  in zero magnetic field. Despite this observation well corresponds to the theoretical predictions, only magnetic field dependence allows to distinguish the non-linear Hall effect from a thermoelectric response. We demonstrate that second-harmonic Hall voltage shows odd-type dependence on the direction of the magnetic field, which is a strong argument in favor of current-magnetization effects. In contrast, one order of magnitude higher thermopower signal is independent of the magnetic field direction.
\end{abstract}

\pacs{73.40.Qv  71.30.+h}

\maketitle

\section{Introduction}

Non-linear Hall effect has been predicted in a wide class of time-reversal invariant materials~\cite{golub,golub1,moore, sodemann,jiang}. In the linear response, there is no Hall current in the presence of time-reversal symmetry. It is argued in Refs.~\onlinecite{golub,golub1,moore, sodemann,jiang}, that a non-linear Hall-like current can arise from the Berry curvature in momentum space. Since Berry curvature often concentrates in regions  where two or more bands cross, three classes of candidate materials have been proposed~\cite{sodemann}: topological crystalline insulators, two-dimensional transition metal dichalcogenides, and three-dimensional Weyl and Dirac semimetals.
Another possible  contribution to non-linear Hall effect is skew scattering with nonmagnetic impurities in time-reversal-invariant noncentrosymmetric materials~\cite{skew}.

Recently,  the time-reversal-invariant non-linear Hall (NLH) effect  has been reported for layered transition metal dichalcogenides~\cite{kang,ma}. It stimulates a search for the Berry curvature dipole induced NLH effect in  three-dimensional  crystals, where  Dirac and Weyl semimetals~\cite{armitage} are excellent candidates, since there is symmetry-protected conic dispersion in the bulk spectrum~\cite{wang1, wang2}. This spectrum has been experimentally confirmed by angle-resolved photoemission spectroscopy (ARPES), e.g., for Cd$_3$As$_2$ Dirac material~\cite{arpes1, arpes2}, and  for MoTe$_2$ and WTe$_2$ type II Weyl semimetals~\cite{armitage,wang,wu}.  Because of low  symmetry, MoTe$_2$ and WTe$_2$ are  advantageous~\cite{sodemann} in a search for the NLH effect. 

\begin{figure}
\centerline{\includegraphics[width=0.8\columnwidth]{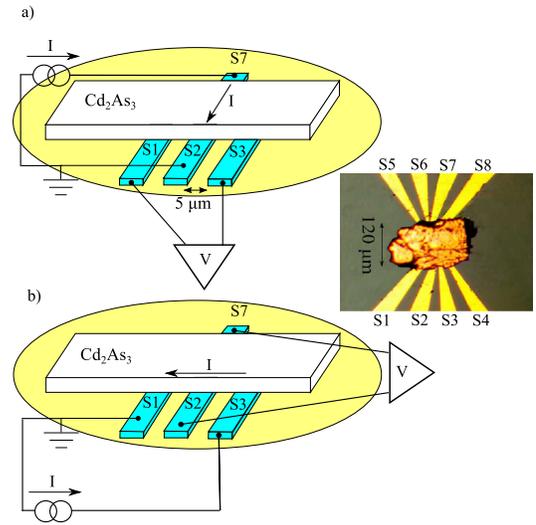}}
\caption{(Color online) Top-view image of the sample with a small Cd$_3$As$_2$ single crystal and the sketch with electrical connections. 100 nm thick and 10 $\mu$m wide Au leads are formed on a SiO$_2$ substrate. A Cd$_3$As$_2$ single crystal ($\approx$ 100~$\mu$m size)  is transferred on top of the leads, forming contacts S1-S8 in regions of $\approx$ 10 $\mu$m overlap between the crystal and the leads. The second-harmonic (2$\omega$) component of the Hall voltage $V$ is investigated in a standard four-point lock-in technique in symmetric (a) and nonsymmetric (b) connection of the Hall voltage probes in respect to the current line (denoted by arrows), which mostly flows along the sample edge between S1 and S3 in the (b) case.}
\label{sample}
\end{figure}

In the experiments~\cite{kang,ma} on two-dimensional WTe$_2$,  the the second-harmonic Hall voltage depends quadratically on the longitudinal current. In the simplified picture, an a.c. excitation current generates sample magnetization, which leads to the  anomalous Hall effect~\cite{ahe} in zero external magnetic field. The latter appears as the second-harmonic Hall voltage, the amplitude is proportional to the square of the bias current. On the other hand, topological materials are  characterized by strong thermoelectric effects~\cite{ptsn,cdas_thermo}, which also appear as a second-harmonic  quadratic signal~\cite{kvon,shashkin}.  For this reason, it is important to experimentally distinguish between the Berry curvature dipole induced NLH effect  and a thermoelectric response while searching for the NLH effect in  nonmagnetic materials.

Here, we experimentally investigate a non-linear Hall effect  for three-dimensional  WTe$_2$ and  Cd$_3$As$_2$ single crystals, representing  Weyl and Dirac  semimetals, respectively. We observe finite second-harmonic Hall voltage, which  depends quadratically on the longitudinal current  in zero magnetic field. Despite this observation well corresponds to the theoretical predictions, only magnetic field dependence allows to distinguish the non-linear Hall effect from a thermoelectric response. We demonstrate that second-harmonic Hall voltage shows odd-type dependence on the direction of the magnetic field, which is a strong argument in favor of current-magnetization effects. In contrast, one order of magnitude higher thermopower signal is independent of the magnetic field direction.

 \begin{figure}
\includegraphics[width=\columnwidth]{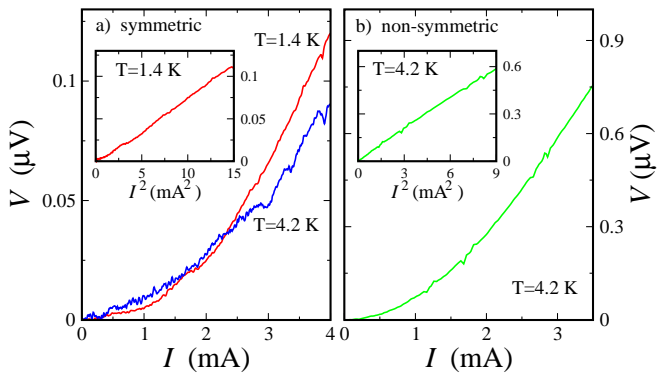}
\caption{(Color online) Examples of $V(I)$ characteristics for a three-dimensional Cd$_3$As$_2$ crystal. Here, $V$ is the second-harmonic (2$\omega$) $xy$ voltage component, $I$ is the ac excitation current at frequency $\omega$. (a)   In the case of the symmetric configuration,  see Fig.~\ref{sample} (a), the measured Hall voltage $V$ is  obviously non-linear, $V\sim I^2$, as it can be seen from the inset. 
The $V(I)$ curve  slightly (about 10\%) depends on temperature in 1.4~K--4.2~K interval.   (b) In the nonsymmetric configuration, depicted in  Fig.~\ref{sample} (b), the  signal level is one order of magnitude higher, about 1~$\mu$V, but the curve is still non-linear $V\sim I^2$, see the inset. The curves are obtained  in zero magnetic field.}
\label{IV}
\end{figure}

\section{Samples and technique}

Cd$_3$As$_2$ crystals were grown by crystallization of molten drops in the convective counterflow of argon held at 5 MPa pressure. For the source of drops the stalagmometer similar to one described~\cite{growth} was applied. About one fifth of the drops were  single crystals. The  energy-dispersive X-ray spectroscopy (EDX) and X-ray powder diffractograms always confirmed pure Cd$_3$As$_2$ with I4$_1cd$ noncentrosymmetric group. 
    
WTe$_2$ compound was synthesized from elements by reaction of metal with tellurium vapor in the sealed silica ampule. The WTe$_2$ crystals were grown by the two-stage iodine transport~\cite{growth1}, that previously was successfully applied~\cite{growth1,growth2} for growth of other metal chalcogenides like NbS$_2$ and CrNb$_3$S$_6$. The WTe$_2$ composition is verified by EDX measurements. The X-ray diffraction  confirms $Pmn2_1$ orthorhombic single crystal WTe$_2$. 

\begin{figure}
\centerline{\includegraphics[width=\columnwidth]{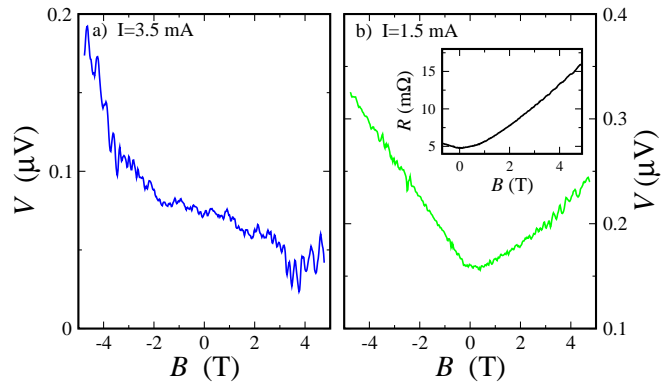}}
\caption{(Color online) Second-harmonic voltage $V$ dependence on the magnetic field $B$ at fixed ac current $I$ for three-dimensional Cd$_3$As$_2$.  (a) In the case of the symmetric voltage probe configuration,   $\Delta V(B)=V(B)-V(B=0)$ is nearly odd function, which is a strong argument in favor of current-magnetization effects. (b) $V(B)$ increases for both field directions  for the nonsymmetric connection scheme,      which allows to identify the thermoelectric response.  Inset demonstrates usual (first-harmonic) Cd$_3$As$_2$ $xx$ magnetoresistance $R(B)$ for our samples. All the curves are obtained at 4.2~K temperature. $I=1.5$~mA is diminishing for (b) with respect to the  $I=3.5$~mA for (a), to avoid overheating effects for high signal in the (b) case.
}
\label{field}
\end{figure}

The initial WTe$_2$ ingot is formed by a large number of small (less than 100 $\mu$m size) WTe$_2$ single crystals, which are weakly connected with each other. In contrast, small  Cd$_3$As$_2$ single crystals  are obtained by a mechanical cleaving method, somewhat similar to described in Ref.~\onlinecite{yu}: we crush the initial 5~mm size Cd$_3$As$_2$ drop onto small fragments. This procedure allows to create a clean Cd$_3$As$_2$ surface  without mechanical polishing or chemical treatment, see Ref.~\onlinecite{cdas} for details.

Fig.~\ref{sample}  shows a top-view image of a sample. The leads pattern is formed by lift-off technique after thermal evaporation of 100 nm Au on the insulating SiO$_2$ substrate. The 10~$\mu$m wide Au leads are  separated by 5~$\mu$m intervals, see  Fig.~\ref{sample}.
Then, a small (about 100 $\mu$m size)   Cd$_3$As$_2$ or WTe$_2$  crystal is transferred to the Au leads pattern and pressed slightly with another oxidized silicon substrate.  A special metallic frame allows to keep substrates parallel and apply a weak pressure to the piece. No external pressure is needed for a crystal to hold on a substrate with Au leads afterward. 

We check by standard magnetoresistance measurements that our Cd$_3$As$_2$ samples demonstrate large magnetoresistance with Shubnikov de Haas oscillations in high magnetic fields~\cite{cdas}. We estimate the concentration of carries as  $n \approx $ 2.3$\times$10$^{18}$~cm$^{-3}$ and low-temperature mobility as $\mu \approx$ 10$^6$~cm$^2$/Vs, which is in the good correspondence with known values~\cite{cdasreview}. Also, we check that our WTe$_2$ samples demonstrate large, non-saturating positive magnetoresistance $(\rho(B)-\rho(B=0))/\rho(B=0)$  in normal magnetic field, which goes to zero in parallel one,  as it has been shown for WTe$_2$ Weyl semimetal~\cite{mazhar}. We do not see Shubnikov de Haas oscillations for WTe$_2$ samples due to lower mobility, see Ref.~\onlinecite{ndwte} for details of magnetoresistance measurements. Examples of the magnetoresistance curves are also shown for our samples in the insets to Figs.~\ref{field} and~\ref{field_wte}.

\begin{figure}
\includegraphics[width=\columnwidth]{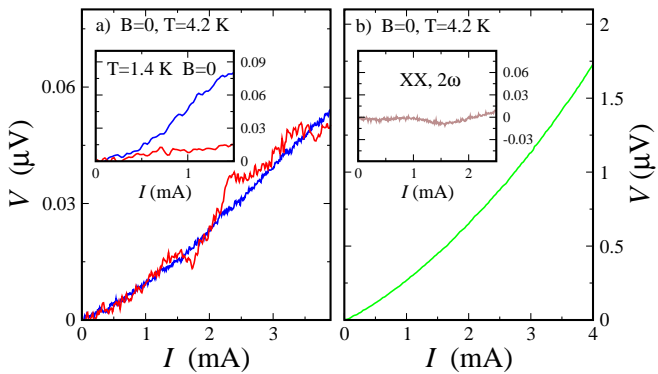}
\caption{(Color online) Examples of $V(I)$ characteristics for a three-dimensional WTe$_2$ crystal. Here, $V$ is the second-harmonic (2$\omega$) $xy$ voltage component, $I$ is the ac excitation current at frequency $\omega$. (a)   In the case of the symmetric configuration,  $V$ is  nonlinear and is similar  for the excitation  current $I$ along both $a$ and $b$ directions at  4.2~K. For lower (1.4~K) temperatures, the Hall voltage  tends to zero for current along the $a$ direction, as depicted in the inset. (b) Nonsymmetric connection leads to high (about 1~$\mu$V) nonlinear $V\sim I^2$ Hall voltage. Inset to (b) demonstrates very small (below 10~nV) second-harmonic $xx$ signal,  i.e. for ac current flowing between S1 and S4, while the potential drop at $2\omega$ is measured between S2 and S3 in Fig.~\ref{sample}. The curves are obtained  in zero magnetic field.
}
\label{IVwte}
\end{figure}

We measure the second-harmonic component of the Hall voltage in standard ac lock-in technique.  To study NLH effect, the special care has been taken to ensure that the Hall probes are symmetric in respect to the current line, as depicted in   Fig.~\ref{sample} (a): the ac current flows between contacts S2 and S7, while the Hall voltage is measured between two neighbour contacts S1 and S3. In this symmetric configuration, there is no temperature gradients between the potential probes, which allows to exclude thermoelectric effects. The latter should be dominant in the nonsymmetric configuration, see  Fig.~\ref{sample} (b), where the current flows between S1 and S3 along the sample edge, while the potential drop is measured across the sample between S2 and S7. 

We ensure, that the measured voltage is antisymmetric with respect to the voltage probe swap and it is independent of the ground probe position. We check, that the lock-in signal is also independent of the modulation frequency (about 110 Hz). The measurements are performed in a standard 1.4~K--4.2~K cryostat equipped with superconducting solenoid.

\section{Experimental results}

Examples of $I-V$ characteristics are shown in Fig.~\ref{IV} for symmetric (a) and nonsymmetric (b) configurations of the voltage probes.   In the case of the symmetric configuration,  like depicted in Fig.~\ref{sample} (a), we obtain clearly non-zero Hall voltage $V^{2\omega}$ for the second harmonics of the ac excitation current $I$. The measured $V^{2\omega}$ is below 0.1~$\mu$V, it slightly (about 10\%) depends on temperature in 1.4~K--4.2~K interval. The $I-V$ curve is obviously non-linear, $V^{2\omega}\sim I^2$, as it can be seen from the inset to Fig.~\ref{IV} (a).

\begin{figure}
\centerline{\includegraphics[width=\columnwidth]{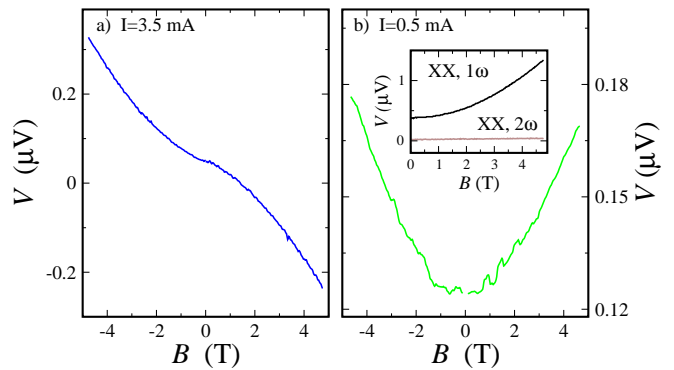}}
\caption{(Color online) Second-harmonic voltage $V$ dependence on the magnetic field $B$ at fixed ac current $I$ for three-dimensional WTe$_2$. (a) In the case of the symmetric voltage probe configuration,   $\Delta V(B)=V(B)-V(B=0)$ is  the odd function of $B$, similarly to the Cd$_3$As$_2$ case. (b) $V(B)$ is nearly even function for the nonsymmetric connection scheme. $I=0.5$~mA is diminishing for (b) with respect to the  $I=3.5$~mA for (a), to avoid overheating effects for high signal in (b) case. Inset to (b) demonstrates usual (first-harmonic) WTe$_2$ $xx$ magnetoresistance for our samples and nearly zero second-harmonic $xx$ component. The curves are obtained at 4.2~K temperature.}
\label{field_wte}
\end{figure}

This behavior well corresponds~\cite{kang,ma} to the expected~\cite{golub,moore, sodemann,jiang} for NLH effect. However, this interpretation can not be accepted without additional arguments. For example, if the potential contacts are not symmetric in respect to the current line, see Fig.~\ref{sample} (b), we also obtain non-linear, $V^{2\omega}\sim I^2$, $I-V$ curve, as presented in Fig.~\ref{IV} (b). In this case the signal level is one order of magnitude higher, about 1~$\mu$V, which better corresponds to typical thermopower values~\cite{cdas_thermo}. 

To experimentally determine the origin of the effect in every of these two cases, we apply an external magnetic field. Fig.~\ref{field} demonstrates second-harmonic voltage dependence on the magnetic field $V^{2\omega}(B)$ at fixed ac current values. In the case of the symmetric configuration, see Fig.~\ref{field} (a),  $\Delta V^{2\omega}(B)=V^{2\omega}(B)-V^{2\omega}(B=0)$ is nearly odd function, i.e. $V^{2\omega}(B)$  depends on the magnetic field direction: $V^{2\omega}(B)$ is diminishing for the positive fields, while it is increasing for the negative ones. In contrast, $V^{2\omega}(B)$ increases for both field directions for the nonsymmetric connection scheme, see Fig.~\ref{field} (b). In this case,  $V^{2\omega}(B)$ even quantitatively resembles Cd$_3$As$_2$ magnetoresistence~\cite{cdas}, which is depicted in the inset to Fig.~\ref{field} (b)  for our samples.

The observed behavior can  be reproduced not only for different samples in different cooling cycles, but also can be demonstrated for another three-dimensional material, like WTe$_2$ Weyl semimetal, see Figs.~\ref{IVwte} and \ref{field_wte}. 

The measured nonlinear second-harmonic Hall voltage $V^{2\omega}$ is also below 0.1~$\mu$V for the symmetric Hall probe connection scheme. Like for Cd$_3$As$_2$ samples, nonsymmetric connection leads to high (about 1~$\mu$V) nonlinear $V^{2\omega}\sim I^2$  voltage, see Fig.~\ref{IVwte} (b).  We also check, that there is no significant second-harmonic signal for the voltage probes situated along the current line, i.e. $xx$ voltage  component is  below 10~nV, see the inset to  Fig.~\ref{IVwte} (b).  

The specifics of NLH effects for layered WTe$_2$ is the strong signal dependence on the current direction~\cite{kang,ma}. In our case of three-dimensional  WTe$_2$, we obtain nearly the same $V^{2\omega}$ for currents along both $a$ and $b$ directions at the liquid helium temperature 4~K, see Fig.~\ref{IVwte} (a). For lower (1.4~K) temperatures, the Hall voltage  $V^{2\omega}$ tends to zero for current along the $a$ direction, as depicted in the inset to  Fig.~\ref{IVwte} (a). This is the  difference of our $I-V$ curves from the layered two-dimensional WTe$_2$, where there was no strong temperature dependence~\cite{kang,ma}.

The similarity between Cd$_3$As$_2$ and WTe$_2$ crystals can also be seen in the magnetic field behavior, see Fig.~\ref{field_wte}. For the symmetric connection scheme, $V^{2\omega}(B)$ demonstrates strong odd-type behavior in respect to the magnetic field direction, as demonstrated in the (a) panel. In contrast, $V^{2\omega}(B)$ is clearly even-type in Fig.~\ref{field_wte} (b), which well corresponds to the bulk WTe$_2$ non-saturating $xx$ magnetoresistance~\cite{mazhar}, see the inset to the panel (b). We wish to note, that there is no noticeable field dependence for the second-harmonic ($2\omega$) $xx$ tensor component, see also the inset to Fig.~\ref{field_wte}.

\section{Discussion}

As a result, we  obtain non-linear second-harmonic $xy$ signal $V^{2\omega}\sim I^2$, which demonstrates different magnetic field behavior, even- or odd-type, for nonsymmetric or strictly symmetric configurations of   voltage probes, respectively.

The odd $V^{2\omega}(B)$ dependence is a good argument for NLH origin of the non-zero second-harmonic Hall voltage: if the ac excitation current  generates sample magnetization, the latter should be sensitive to the direction of  external magnetic field. More precisely, it is possible to demonstrate~\cite{kinetic} from the kinetic equation (in the spirit of Ref.~\onlinecite{sodemann}), that second - order response is absent in classical Hall effect, while it is an odd function of magnetic field for the spectrum with Berry curvature (Weyl semimetals).  

 In contrast, thermoelectric effects are defined by the sample heating, which is proportional to $RI^2$ in our case, i.e. they also produce the second-harmonic response. The magnetic field dependence should be mainly defined by the magnetoresistence $R(B)$, since it is extremely  strong in Weyl and Dirac semimetals. Thus, the thermoelectric response can  not be sensitive to the magnetic field direction. In the experiment, $V^{2\omega}(B)$ even quantitatively resembles $R(B)$ magnetoresistence, see Figs.~\ref{field} and~\ref{field_wte} for our samples.  Note, that Nernst effect can not contribute to the measured $xy$ voltage, since the temperature gradient is along the $y$ axis  in the geometry of the experiment. On the other hand, the Seebeck effect is also characterized~\cite{steele} by even, $R(B)$-like magnetic field dependence.

Thus, we can identify high second-harmonic signal as thermoelectric voltage for nonsymmetric connection schemes, while low $V^{2\omega}$ reflects NLH effect for the strictly symmetric ones.

 For both  connection schemes, some admixture of the effects is possible. We can not completely avoid an asymmetry of the potential contacts, so an admixture of $R(B)$ produces distortions in high fields in Fig.~\ref{field} (a).  On the other hand, NLH effect should be present also in 
the nonsymmetric connection scheme, where the thermoelectric response dominates. Due to the strong odd field dependence of NLH voltage, it can be responsible for the observed $V^{2\omega}$ branch asymmetry in  Figs.~\ref{field} (b) and~\ref{field_wte} (b).

While NLH effect was originally proposed~\cite{sodemann} for Weyl and Dirac semimetals, it can only be seen for noncentrosymmetric crystals. This requirement is obviously fulfilled for the type II Weyl semimetal WTe$_2$, but there is a discussion in the case of Cd$_3$As$_2$.  Ref.~\onlinecite{cdas_sym} insists on the  near-centrosymmetric structure with the space group I4$_1/acd$. On the other hand, the previously established~\cite{cdas1968} I4$_1cd$ noncentrosymmetric group is also confirmed in recent investigations~\cite{cdas_nonsym} and this crystal symmetry is in a reasonable correspondence with ARPES data on the Cd$_3$As$_2$ electronic structure~\cite{cdas_nonsym}. This difference should originate from the Cd$_3$As$_2$ growth method, e.g. the X-ray diffraction  confirms I4$_1cd$ noncentrosymmetric group for our samples. Also, in our case, strain may occur at SiO$_2$-Cd$_3$As$_2$ interface due to materials misfit, which affects the initial symmetry~\cite{strain}.
It is worth mentioning, that skew scattering is allowed in all noncentrosymmetric crystals, whereas Berry curvature dipole requires more strict symmetry conditions~\cite{skew}. We still can not distinguish  these two contributions to NLH effect in the present experiment.

\section{Conclusion}
We experimentally investigate a non-linear Hall effect  for three-dimensional  WTe$_2$ and  Cd$_3$As$_2$ single crystals, representing  Weyl and Dirac  semimetals, respectively. We demonstrate finite second-harmonic Hall voltage, which  depends quadratically on the longitudinal current  in zero magnetic field.  If the potential contact are perfectly symmetric in respect to the current line, the observed signal is in the nanovolt range. It shows odd-type dependence on the direction of the magnetic field, which is a strong argument in favor of current-magnetization effects. If the potential contact are strongly nonsymmetric, temperature gradient produces one order of magnitude higher thermopower signal with even-type magnetic field dependence.

\acknowledgments
We wish to thank Leonid E. Golub, Yu.S. Barash, and V.T.~Dolgopolov for fruitful discussions.  We gratefully acknowledge financial support partially by the RFBR  (project No.~19-02-00203), RAS, and RF State contracts.


\begin{thebibliography}{99}
\bibitem{kinetic} The calculation will be published elsewhere.
\end{thebibliography}
\end{document}